\documentclass[usenatbib]{mn2e}
\usepackage{psfig}

\newcommand{\etal}{et~al.}
\newcommand{\ionhy}{H{\sc ii}}
\newcommand{\UCHII}{UCH{\sc ii}}

\newcommand{\ammonia}{$\mbox{NH}_{3}$}

\newcommand{\vsfig}[2]           
{
  \begin{center}
    \begin{minipage}[t]{0.05\textwidth}
      {\footnotesize \raisebox{40mm}{(#2)}}
    \end{minipage}
    \begin{minipage}[t]{0.42\textwidth}
      \psfig{file=./#1.ps,height=0.95\textwidth,angle=270}
    \end{minipage}
    \hfill
  \end{center}
}

\begin{document}

\title[Extended emission from young \ionhy\ regions] 
{Extended emission associated with young \ionhy\ regions}

\author[Ellingsen, Shabala \& Kurtz]{S.P. Ellingsen$^1$, S.S. Shabala$^1$, 
  S.E. Kurtz$^2$\\
$^1$ School of Mathematics and Physics, University of Tasmania, 
Private Bag 21, Hobart, Tasmania 7001, Australia;\\
Simon.Ellingsen@utas.edu.au\\
$^2$ Centro de Radioastronom\'ia y Astrof\'isica, UNAM-Morelia, 
Adpo. Postal 3-72, C.P. 58089,
Morelia, Michoac\'an, M\'exico}

\maketitle

\begin{abstract}
  
  We have used the Australia Telescope Compact Array (ATCA) to make
  observations of a sample of eight young ultra-compact \ionhy\/
  regions, selected on the basis that they have associated class~II
  methanol maser emission.  We have made observations sensitive to
  both compact and extended structures and find both to be present in
  most sources.  The scale of the extended emission in our sample is
  in general less than that observed towards samples based on {\em
    IRAS} properties, or large single-dish flux densities.  Our
  observations are consistent with a scenario where extended and
  compact radio continuum emission coexists within \ionhy\ regions for
  a significant period of time.
  
  We suggest that these observations are consistent with a model where
  \ionhy\/ evolution takes place within hierarchically structured
  molecular clouds.  This model is the subject of a companion paper
  \citep{SEKF04} and addresses both the association between compact
  and extended emission and \UCHII\/ region lifetime problem.

\end{abstract}


\section{Introduction}

Ultra-compact \ionhy\ regions (\UCHII) arguably represent the earliest
stage of high-mass star formation that can be reliably identified.  As
such they have been the focus of intensive investigation, particularly
since the first large scale interferometric survey of this class of
sources by \citet{WC89b}.  This survey detected compact radio continuum
emission towards two-thirds of the sources searched, a much higher
fraction than expected.  Simple modelling of \ionhy\/ regions
\citep[e.g.][]{K54} suggests that the ultra-compact phase should be
short lived and hence seen in only a small fraction of sources.  This
discrepancy (known as the \UCHII\ lifetime problem), is thought to be
due to the presence of an additional confinement mechanism which
partly counteracts the expected pressure driven expansion.  Numerous
mechanisms have been suggested to reduce the expansion rate of
\ionhy\/ regions, either through confinement due to infall
\citep{RHBMJS80}, bow shocks \citep{VMWC90}, dense warm environments
\citep*{DRG95}, turbulent pressure \citep{XMVH96}, or other means such
as photoevapouration of circumstellar discs \citep{HJLS94} and
mass-loaded stellar winds \citep*{DWR95}.  Champagne flows \citep{T79}
have also been invoked as a means of extending the lifetime of the
ultra-compact phase of \ionhy\/ regions.  Although each model appears
able to explain some objects, none seem to be universally valid. The
\UCHII\ lifetime issue has been most recently reviewed by
\citet{KCCHW00}.

A second observational challenge related to \UCHII\ regions has
recently emerged from their apparent association with extended radio
continuum emission.  The majority of interferometric observations of
\UCHII\ regions have been sensitive to emission on angular scales of
the order of 1-10 arcseconds.  However, surveys by \citet{KWHO99} and
\citet{KK01} have shown the presence of significant emission on larger
scales towards the majority of the \UCHII\ regions observed.
Morphologically the extended emission appears to be directly
associated with the \UCHII\ regions (rather than a projection effect)
and this is supported by observations of radio recombination lines
\citep{KK01}.  The coexistence of a compact, high-emission measure
region within a larger diffuse, lower-emission measure zone is not
predicted by models which solve the lifetime problem through
confinement.  It may be broadly consistent with replenishment
solutions, such as photoevapouration of circumstellar disks
\citep{HJLS94}, but no detailed work to predict \ionhy\ region
morphology has been undertaken for this model.

A scenario which addresses both the lifetime problem and the
extended emission problem is the following :
\begin{enumerate}
\item A high-mass star commences nuclear fusion in the core and starts
  ionizing the surrounding neutral material, rapidly forming a \UCHII\
  region.
\item Soon after, a zone of more diffuse ionized gas begins to form
  surrounding the \UCHII\ region.  Over time the size of the diffuse
  region grows and the compact region begins to dissipate.
\item Eventually the compact region dissipates entirely and we are left
  with a classical \ionhy\ region.
\end{enumerate}
There may be some \ionhy\/ regions that do not follow this scenario,
such as those associated with later type stars that produce relatively
few ionising photons, where it is possible that no diffuse ionized
region forms \citep{SEKF04}.  However, here we are interested in the
large fraction of \ionhy\/ regions that have been observed to exhibit
both compact and more diffuse emission.

We will leave aside for the moment the question as to how this
scenario occurs, this is addressed in more detail in
section~\ref{sec:discussion} and \citet{SEKF04}.  Our scenario is
consistent with current observations, as any observation made with an
interferometer at high resolution during stage (ii) will detect
compact emission and observations made with lower resolution will
detected extended emission.  All we require to solve the dual
observational challenges of lifetime and extended emission is that the
timescale over which compact and extended emission coexist is long
enough to explain the observed excess of \UCHII\ regions.  On the
basis of various arguments given by \citet{KCCHW00} this is estimated
to be approximately a factor of 5 (although uncertain by a factor of
2) and so using $3 \times 10^4$ years as an estimate for the lifetime
of the \UCHII\ phase \citep{WC89a}, we estimate stage (ii) to be of
the order of $10^5$ years.

If the scenario outlined above is correct then the flux density and
radius of the diffuse extended emission will increase with time.  Very
young \ionhy\/ regions will have little or no associated extended
emission, while older regions will have a significant amount.  Masers
are thought to trace the early stages of high-mass star formation and
so we would predict that \ionhy\/ regions with associated masers
should show little or no extended emission.  Class~II (e.g. 6.7-GHz)
methanol masers are believed to exclusively trace the early stages of
high-mass star formation \citep{MENB03}.  This contrasts with the
other common masing molecules OH and water which are associated with
more than one type of astrophysical object.  Many 6.7-GHz methanol
masers have no associated centimetre radio continuum emission
\citep{PNEM98,WBHR98}, but are associated with millimetre and
sub-millimetre continuum emission \citep*{PHB02,WMABL03}.  This
suggests that many class~II methanol masers trace a pre-\UCHII\ phase,
and for those where there is an associated \UCHII\ region it is young.
Analysis of the scale-height of 6.7-GHz methanol masers in the Galaxy
shows that it is significantly smaller than any other extreme
Population I objects \citep{VRGM96}.  

To test the scenario outlined above we have selected a sample of eight
\UCHII\ regions associated with 6.7-GHz methanol masers.  Assuming
that this criteria selects young \ionhy\/ regions then we predict they
should exhibit relatively little extended emission compared with the
regions observed by \citet{KWHO99} and \citet{KK01}.

\section{Observations \& Data processing}

Eight \UCHII\/ regions associated with 6.7-GHz methanol masers were
imaged with the Australia Telescope Compact Array (ATCA) in the 750D
configuration.  For the 750D array the minimum baseline length is 107
m and the maximum is 719 m.  The observations were made on 1999 July
10\&11, with all sources being observed on both days to improve the
overall $uv$ coverage.  For each of the \UCHII\ regions a 3-minute
scan was both preceded and followed by a 1-minute scan of a phase
calibrator.  Over the two days sources were observed approximately 20
times, for a total onsource integration time of approximately 1 hour.
The correlator was configured to record a 128-MHz bandwidth, centred
at a frequency of 8.64 GHz.  The data were calibrated using the {\tt
  miriad} software package applying the standard techniques for ATCA
continuum observations.  The data for each day were calibrated
separately and merged into a single dataset for imaging.
Table~\ref{tab:fields} lists the fields imaged, the RMS level in the
residual image and similar information for the related 6-km array
observations (see below).

\begin{table*}
  \caption{The fields imaged in the 750D array with the ATCA.  The listed
  {\em IRAS} sources are all within 30 arcsec of the 6.7-GHz methanol
  maser position, except for G\,318.95-0.20, where the separation is 144 
  arcsec.}
  \begin{tabular}{lccrlrl} \hline
    {\bf Field} & {\bf Right Ascension} & {\bf Declination} & 
      \multicolumn{1}{c}{\bf RMS in} & {\bf 6km Obs.}  & 
      \multicolumn{1}{c}{\bf RMS in 6km} & {\bf Associated} \\
    {\bf Name}  & {\bf (J2000)}         & {\bf (J2000)}     & 
      \multicolumn{1}{c}{\bf resid. image} & {\bf Reference} & 
      \multicolumn{1}{c}{\bf resid. image} & {\bf IRAS source} \\ 
                &                       &                   &
      \multicolumn{1}{c}{\bf (mJy beam${-1}$)} &             & 
      \multicolumn{1}{c}{\bf (mJy beam${-1}$)} & \\ \hline
    G\,308.92+0.12 & 13:43:02 & -62:08:51 &  0.1 & \citet{PNEM98} &  0.2 & 
    13395-6153 \\ 
    G\,309.92+0.48 & 13:50:42 & -61:35:10 &  0.2 & \citet{PNEM98} &  0.3 & 
    13471-6120 \\ 
    G\,318.95-0.20 & 15:00:55 & -58:58:42 &  1.0 & this work      &  0.2 & 
    14567-5846 \\ 
    G\,328.81+0.63 & 15:55:48 & -52:43:07 &  0.5 & this work      &  0.5 & 
    15520-5234 \\ 
    G\,336.40-0.25 & 16:34:11 & -48:06:26 &  0.2 & \citet{PNEM98} &  1.5 & 
               \\ 
    G\,339.88-1.26 & 16:52:05 & -46:08:34 & 0.08 & \citet{ENM96}  & 0.15 & 
    16484-4603 \\ 
    G\,345.01+1.79 & 16:56:48 & -40:14:26 &  0.2 & this work      &  0.5 & 
    16533-4009 \\ 
    NGC6334F       & 17:20:53 & -35:47:01 &  2.2 & \citet{ENM96}  & 59.0 & 
    17175-3544 \\ \hline 
  \end{tabular}
  \label{tab:fields}
\end{table*}

Imaging and self-calibration of the data was undertaken in {\tt
  difmap}.  To identify all sources of emission within the primary
beam a 2048x2048 arcsecond image, centred at the phase centre was
created and cleaned.  The {\tt difmap} model was then discarded and
the image re-cleaned and self-calibrated with a small loop gain and
clean boxes around all emission regions.  The amplitude corrections
applied by self-calibration were typically small (a few percent or
less), indicating good basic calibration of the data.  The only
exceptions were the few sources that exhibited significant extended
emission that was not well sampled with the 750D array.  For the
majority of the sources the resulting RMS after imaging was 1 mJy
beam$^{-1}$ or less.

High resolution ATCA observations have been published in the
literature for G\,318.91-0.16, G\,339.88-1.26, NGC6334F \citep*{ENM96}
and G\,308.92+0.12, G\,309.92+0.48, G\,336.41-0.26 \citep{PNEM98}.
The imaging methodology used for the 750D observations was based on
that of \citet{PNEM98} and so for consistency the sources observed by
\citet{ENM96} were re-imaged using the same approach.  The remaining
two \UCHII\/ regions (G\,328.81+0.63 and G\,345.01+1.79) had not
previously been imaged at high resolution and sensitivity.  These
sources were observed at 8.59~GHz with the ATCA in the 6A
configuration on 1994 July 4 (G\,328.81+0.63) and July 5
(G\,345.01+1.79).  These data were calibrated using the
$\mathcal{AIPS}$ software package, applying the standard techniques
for ATCA continuum observations.  Imaging was undertaken in {\tt
  difmap}, using the same approach as for the 750D observations.
PKS1934-638 was used as the primary flux density calibrator for all
observations, its flux density at 8.59 and 8.64 GHz was assumed to be
2.86 and 2.84 Jy respectively.

\section{Results}

Figures~\ref{fig:g308}-\ref{fig:ngc6334f} show contour plots of radio
continuum images for each of the eight fields (containing a total of
11 \ionhy\ regions).  In each figure (except Fig.~\ref{fig:g345})
panel {\em a} shows a 50 x 50 arcsec 8.6-GHz image made using a 6-km
array, panel {\em b} shows a 5 x 5 arcmin 8.6-GHz image obtained using
a 750-m array and panel {\em c} shows a 10 x 10 arcmin 843-MHz MOST
image from the Molonglo Galactic Plane Survey (MGPS1) \citep{GCLY99}
for sources where they are available.  The location of 6.7-GHz
methanol maser clusters is marked with a plus symbol.  The maximum
angular resolution of the 6-km array observations is approximately 1
arcsec and the largest angular scales that can reliably be imaged is
approximately 15 arcsec.  For the 750-m array observations the angular
resolution is approximately 7 arcsec and the largest angular scale
that can be detected is of the order of 50 arcsec.  The MGPS1 images
have an angular resolution of 43 arcsec and the RMS noise is typically
1-2 mJy~beam$^{-1}$.  The peak and integrated flux density for the
ATCA observations are listed in Table~\ref{tab:dynamic}.  These
observations are at the same frequency and have comparable angular
scales and sensitivities to the VLA B- and D-array observations of
\citet{KWHO99}.  In contrast to the {\em IRAS} selected \ionhy\ 
regions of \citeauthor{KWHO99} where many of the D-array images
(equivalent to the 750-m array images) show significant extended
emission, for our sample that is not the case.  The only source which
shows significant large scale structure is G\,336.41-0.26 and in this
source the compact region lies at the edge of the diffuse emission and
so is probably not directly associated.
 
Our sample was selected on the basis of having a detected \UCHII\ 
region with an associated 6.7-GHz methanol maser site, the only
exception being G\,318.95-0.20 for which \citet{ENM96} detected no
radio continuum emission at the maser site, but did detect an \ionhy\ 
(G\,318.91-0.16) region associated with {\em IRAS}\ 14567-5846 which
is 2 arcmin away.  Our 750-m array observations are consistent with
the higher resolution observations of \citet{ENM96}, detecting radio
continuum emission only at the {\em IRAS} site and not at the maser
location.  Three additional regions of radio continuum were detected
which are not associated with methanol maser emission.  G\,345.00+1.79
and G\,345.01+1.82 can clearly be see in Fig.~\ref{fig:g345} and NGC6334E
is present to the north-west of NGC6334F in Fig.~\ref{fig:ngc6334f}b
and dominates the emission in Fig.~\ref{fig:ngc6334f}c.

\subsection{Individual sources}

{\em G\,308.92+0.12:} This \ionhy\ region has a core-halo morphology
and in terms of the size of the emission region seen in the 6-km array
image (Fig.~\ref{fig:g308}a) it is one of the largest associated with
a maser in this sample.  \citet{PNEM98} found the 6.7-GHz methanol
masers are located near the northern tip of the region, offset by
approximately 5 arcsec from the peak of the radio continuum emission
and suggest that it is likely there is more than one ionizing source
present.  This is supported by mid-infrared observations which
detected a strong 11.5-\micron\ point source associated with the maser
location, but nothing at the \ionhy\ region peak \citep{PSENM04}.  The
750-m array image (Fig.~\ref{fig:g308}b) shows that the true extent of
the \ionhy\ region is significantly larger than suggested by the
observations of \citet{PNEM98} and slight extensions are present to
the north-west, east and south.  The 750D observations show that the
flux density is increasing with decreasing baseline length, suggesting
there may be more flux density present on still larger scales.  The
MOST image (Fig.~\ref{fig:g308}c) lends some support to this showing
possible extension to the north-west.

\begin{figure}
\vsfig{g308.92_h}{a}
\vsfig{g308.92_m}{b}
\vsfig{g308.92_l}{c}
\caption{Radio continuum images of G\,308.92+0.12 (a) 8.59-GHz ATCA 6-km image 
  of the compact emission \citep{PNEM98} (b) 8.64-GHz ATCA 750-m image
  of the extended emission (c) 843-MHz MOST image of the large scale
  continuum emission \citep{GCLY99}.  The location of the
  6.7-GHz methanol maser cluster is marked with a plus symbol.  The
  squares in (b) and (c) show the area covered in figures (a) and (b)
  respectively.}
\label{fig:g308}
\end{figure}

{\em G\,309.92+0.48:} This \ionhy\ region ({\em IRAS}\ 13471-6120) is
the most compact in the sample with Table~\ref{tab:dynamic} showing
only 8 per cent more integrated flux density in the 750-m array
observations than in the 6-km, the lowest ratio observed.  The 6-km
array image (Fig.~\ref{fig:g309}a) of \citet{PNEM98} shows slight
low-level extension of the main feature with a second region with a
peak flux density of 5.4 mJy beam$^{-1}$ to the north-east.
Observations at 10.8 and 18.2\micron\ by \citet*{DPT00} detect strong
emission associated with the \ionhy\ region and a weaker more deeply
embedded source associated with the weak radio continuum emission to
the north-east.   \citet{PNEM98} found the 6.7-GHz methanol masers to
be located close to the peak of the \ionhy\ region.  The 750-m array
image (Fig.~\ref{fig:g309}b) also shows small deviations from a point
source with lower level contours slightly extended to the north-east,
south-east and west.

\begin{figure}
\vsfig{g309.92_h}{a}
\vsfig{g309.92_m}{b}
\vsfig{g309.92_l}{c}
\caption{Radio continuum images of G\,309.92+0.48 (a) 8.64-GHz ATCA 6-km 
  image of the compact emission \citep{PNEM98} (b) 8.64-GHz ATCA 750-m
  image of the extended emission (c) 843-MHz MOST image of the large
  scale radio continuum emission \citep{GCLY99}. The location of the
  6.7-GHz methanol maser cluster is marked with a plus symbol.  The
  squares in (b) and (c) show the area covered in figures (a) and (b)
  respectively.}
\label{fig:g309}
\end{figure}

\begin{figure}
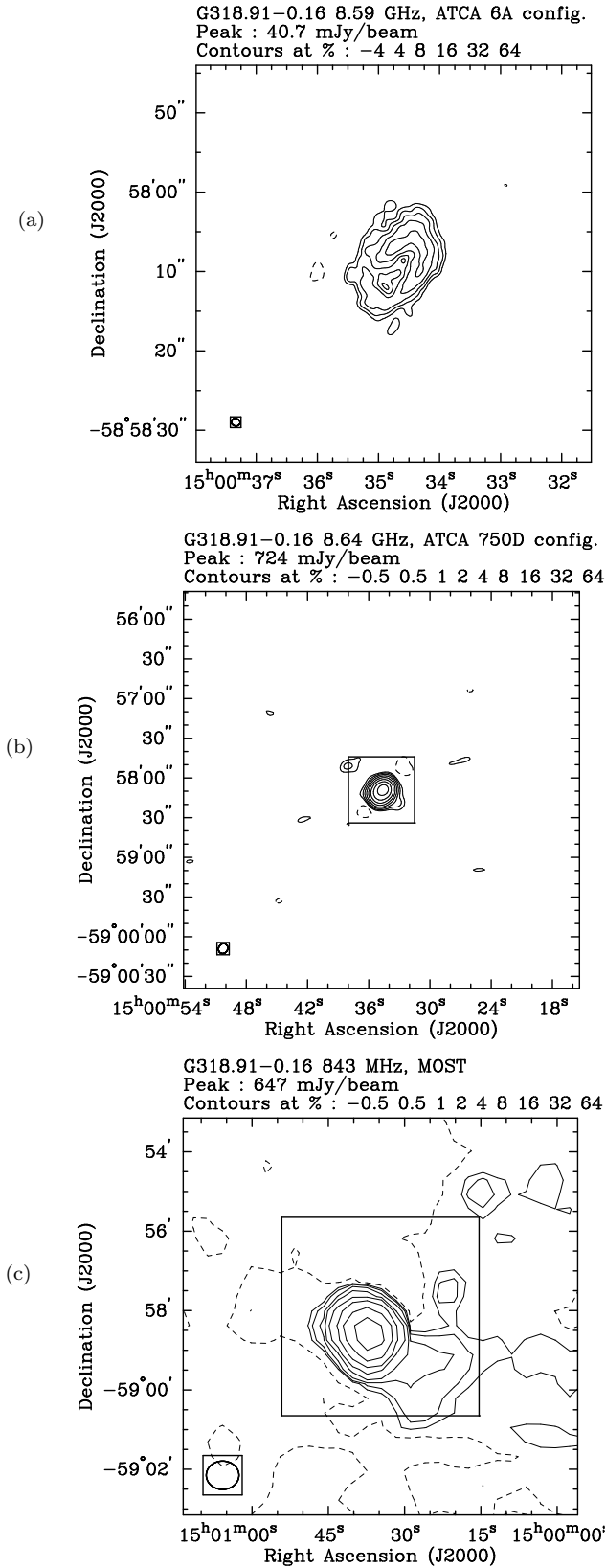

\vsfig{g318.91_h}{a}
\vsfig{g318.91_m}{b}
\vsfig{g318.91_l}{c}
\caption{Radio continuum images of G\,318.91-0.16 (a) 8.59-GHz ATCA 6-km 
  image of the compact emission (b) 8.64-GHz ATCA 750-m image of the
  extended emission (c) 843-MHz MOST image of the large scale radio
  continuum emission \citep{GCLY99}. The squares in (b) and (c) show
  the area covered in figures (a) and (b) respectively.}
\label{fig:g318}
\end{figure}

{\em G\,318.91-0.16:} The \ionhy\ region shown in Fig.~\ref{fig:g318}
({\em IRAS}\ 14567-5846) has a shell morphology at the highest
resolutions.  There is no compact radio continuum emission associated
with the masers to a 5-$\sigma$ level of 0.82 mJy beam$^{-1}$
\citep{ENM96} and we did not detect any diffuse emission with a
5-$\sigma$ limit of 5 mJy beam$^{-1}$.  The \ionhy\ region is quite
symmetrical in the 750-m array image (Fig.~\ref{fig:g318}b), but shows
increasing flux density with decreasing baseline length.  This,
combined with low-level extension to the south-west (also present in
the MOST image Fig.~\ref{fig:g318}c) suggests further extended
emission is present on still larger scales.

{\em G\,328.81+0.63:} The \ionhy\ region ({\em IRAS}\ 15520-5234)
shown in Fig.~\ref{fig:g328}a appears to be a partial superposition of
two nearby sources, a point source and a cometary \ionhy\ region.  Our
image is consistent with that of \citet{WBHR98}, but our higher
dynamic range better reveals the cometary morphology.
\citeauthor{WBHR98} found the majority of 6.7-GHz methanol masers to
be located near the peak of the cometary region, but offset slightly
in the direction of the unresolved peak.  There is a single maser spot
offset by approximately 3 arcsec from the others, close to the peak of
the unresolved continuum emission \citep*{NWCWG93,DOE04}.
Mid-infrared observations by \citet{DPT00} detect at least 6 sources
in this region and morphological comparison with the radio continuum
emission suggests that the masers are coincident with the peak of the
infrared emission.  The 6-km image shows continuous low-level
extensions to the north and south, and regions of emission further
north and south are suggestive of large scale extensions not
satisfactorily imaged with this array configuration.  The 750m array
image shows slightly resolved, nearly symmetrical emission with the
flux density near constant on baselines shorter than 30k$\lambda$.

\begin{figure}
\vsfig{g328.81_h}{a}
\vsfig{g328.81_m}{b}
\vsfig{g328.81_l}{c}
\caption{Radio continuum images of G\,328.81+0.63 (a) 8.59-GHz ATCA 6-km image
  of the compact emission (b) 8.64-GHz ATCA 750-m image of the
  extended emission (c) 843-MHz MOST image of the large scale radio
  continuum emission \citep{GCLY99}.  The location of the 6.7-GHz
  methanol maser cluster is marked with a plus symbol.  The squares in
  (b) and (c) show the area covered in figures (a) and (b)
  respectively.}
\label{fig:g328}
\end{figure}

{\em G\,336.41-0.26:} There is a large amount of extended emission in
the vicinity of this \ionhy\ region and this is responsible for the
poor dynamic range of both the 6-km and 750-m array images.
\citet{PNEM98} report four sites of 6.7-GHz methanol maser emission in
this field, but none are directly associated with the strongest
compact radio continuum emission in the region.  Single featured, weak
masers are associated with the continuum emission in
Fig.~\ref{fig:g336}a immediately to the west (G\,336.409-0.257) and
north-west (G\,336.410-0.256) of the \ionhy\ region.
Fig.~\ref{fig:g336}b and \ref{fig:g336}c show a ridge of extended
emission running south-west to north-east with the four maser sites
lying along a line with the same position angle as the southern edge
of the ridge.  Interestingly there is no compact emission associated
with the other two maser sites (G\,336.404-0.254 \& G\,336.433-0.262)
each of which contains multiple components.  \citet{DPT00} didn't
detect any mid-infrared emission associated with the G\,336.433-0.262
masers, but the other masers and compact radio continuum peak were not
within the field of view of their observations.  For this source it
would appear that the extended emission is not directly associated
with the \UCHII\ region, as it lies at the edge, rather than near the
centre.  MSX images of the same region in the E-band (21~\micron) show
a similar morphology to the MOST image.  We suggest that something has
triggered a new epoch of star formation that we see projected along
the southern edge of radio continuum emission associated with a
previous episode.  If this is the case then there are likely to be are
a number of millimetre continuum sources along this southern ridge
(corresponding to the maser locations) that will in time produce
detectable \ionhy\ regions.

\begin{figure}
\vsfig{g336.41_h}{a}
\vsfig{g336.41_m}{b}
\vsfig{g336.41_l}{c}
\caption{Radio continuum images of G\,336.41-0.26 (a) 8.64-GHz ATCA 6-km image
  of the compact emission \citep{PNEM98} (b) 8.64-GHz ATCA 750-m image
  of the extended emission (c) 843-MHz MOST image of the large scale
  continuum emission \citep{GCLY99}.  The location of the 6.7-GHz
  methanol maser clusters are marked with plus symbols.  The squares in
  (b) and (c) show the area covered in figures (a) and (b)
  respectively.}
\label{fig:g336}
\end{figure}

{\em G\,339.88-1.26:} This \ionhy\ region ({\em IRAS}\ 16484-4603) is
relatively weak and the 6-km image (Fig.~\ref{fig:g339}a) shows signs
of low-level extended emission, particularly to the north-east and
north-west.  The image in Fig.~\ref{fig:g339}a is based on the same
calibrated data set as that presented by \citet{ENM96}, but has been
cleaned and self-calibrated using {\tt difmap} for consistency with
the treatment applied to the other sources in the sample.
\citet{ENM96} estimate G\,339.88-1.26 to be produced by a B0.5 star.
High resolution mid-infrared observations by \citet{DWPPT02} show
three sources near the \ionhy\ region, elongated perpendicular to the
radio continuum emission.  The relative alignment of radio and
mid-infrared emission is not certain, with that proposed by
\citeauthor{DWPPT02} being predicated on the argument that the
\ionhy\ region is associated an star visible at optical and
near-infrared wavelengths located in front of the mid-infrared source.
A counter argument to this is that there should be no compact radio
continuum emission, nor 6.7-GHz methanol masers remaining if the
exciting star is optically visible.  The 750-m array image
(Fig.~\ref{fig:g339}b) shows significant extended emission and the
additional low-level features surrounding the main region suggests
that there is additional flux on still larger scales.

\begin{figure}
\vsfig{g339.88_h}{a}
\vsfig{g339.88_m}{b}
\vsfig{g339.88_l}{c}
\caption{Radio continuum images of G\,339.88-1.26 (a) 8.59-GHz ATCA 6-km image
  of the compact emission \citep{ENM96} (b) 8.64-GHz ATCA 750-m image
  of the extended emission (c) 843-MHz MOST image of the large scale
  continuum emission \citep{GCLY99}.  The location of the 6.7-GHz
  methanol maser cluster is marked with a plus symbol.  The squares in
  (b) and (c) show the area covered in figures (a) and (b)
  respectively.}
\label{fig:g339}
\end{figure}

\begin{figure}
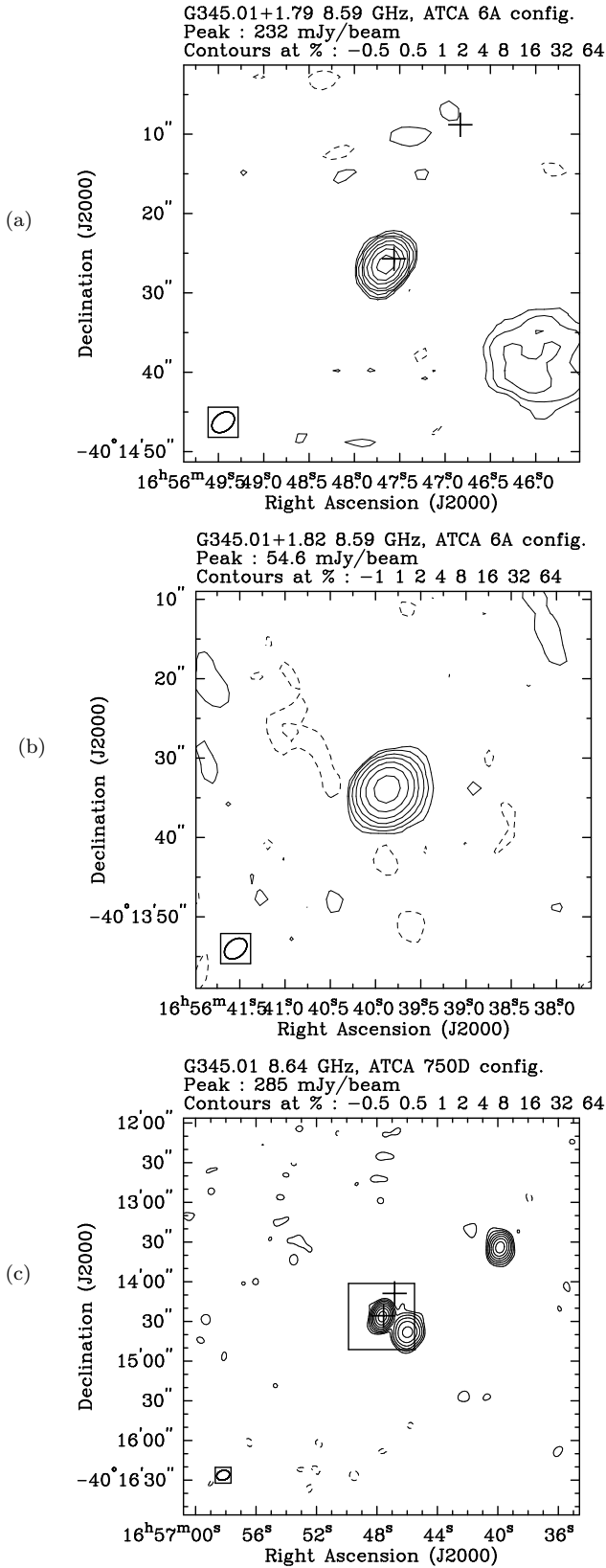

\vsfig{g345.01_h}{a}
\vsfig{g345.01_h2}{b}
\vsfig{g345.01_m}{c}
\caption{Radio continuum images of G\,345.01+1.79 (a) 8.59-GHz ATCA 6-km image
  of the compact emission (b) 8.59-GHz ATCA 6-km image of the compact
  emission from the \ionhy\ region G\,345.01+1.82 (c) 8.64-GHz ATCA
  image of the extended emission from the G\,345.01+1.79 region.  The
  location of the 6.7-GHz methanol maser clusters are marked with plus
  symbols.  The square in (c) shows the area covered in figure (a).}
\label{fig:g345}
\end{figure}

{\em G\,345.01+1.79} This \ionhy\ region ({\em IRAS}\ 16533-4009) is
very compact with less than 10 per cent more integrated flux density
in the 750-m array observations than for the 6-km.  \citet{WBHR98}
found that the 6.7-GHz methanol masers are offset to the west of the
centre of the \ionhy\ region.  G\,345.01+1.79 holds the distinction of
exhibiting methanol maser emission in more transitions than any other
known source \citep{V98,VESKOV99,CSECGSD01}.  The mid-infrared
emission is also compact, being perhaps marginally resolved at
18.2\micron\ \citep{DPT00}.  There are two additional radio continuum
sources in the same field of view, a lower surface brightness region
to the south-west of about 12 arcsec in size (G\,345.00+1.79) and a
compact region to the north-west (G\,345.01+1.82) shown in
Fig.~\ref{fig:g345}b.  There is no MGPS1 image for this source as it
lies more than 1.5 degrees from the Galactic Plane.

{\em NGC6334F:} This cometary \ionhy\ region ({\em IRAS}\ 17175-3544)
lies in a star formation region that has been extensively studied at
radio \citep*{RCM82}, submillimetre \citep{S00}, near infrared
\citep*{SHM89}, mid infrared \citep{KDJHFHD99}, and x-ray
\citep{SMKTYU00} wavelengths and in molecular lines
\citep{KJ99,MSMKSDS00}.  There are six \ionhy\ regions in the NGC6334
region of which F (Fig.~\ref{fig:ngc6334f}a) is the most compact
\citep{RCM82}.  The profusion of sources in, and the large number of
observations of the NGC6334 region has lead to a complex and confusing
nomenclature which is detailed in the appendix of \citet{KJ99}.  The
NGC6334F \ionhy\ region is referred to as NGC6334I in infrared
observations and here we have used the nomenclature appropriate to
each wavelength.

NGC6334 provides nice a demonstration as to why class~II methanol
masers are thought to trace only the early stages of high-mass star
formation.  There are four sites of 6.7-GHz methanol masers in the
NGC6334 region, two are close to NGC6334F (Fig.~\ref{fig:ngc6334f}a),
with one projected against the leading edge of the \ionhy\ region and
the second offset to the north-west 6 arcseconds \citep{ENM96}.  The
third 6.7-GHz methanol maser G\,351.54+0.66 (Fig.~\ref{fig:ngc6334f}b)
is associated with the high-mass class~0 candidate NGC6334I(N),
although offset from the peak of the dust continuum emission
\citep{S00}.  The final 6.7-GHz methanol maser G\,351.16+0.70 is
associated with the NGC6334V region which also has water and ammonia
maser emission.  The NGC6334 region also contains three sites of OH
maser emission \citep{BW01}.  The strongest of these is coincident
with the methanol masers in NGC6334F, while NGC6334V and the more
evolved regions NGC6334A also have OH masers.  The two class~II
methanol maser sites without associated OH maser emission both appear
to be young star forming regions.  NGC6334I(N) is cold, dense,
optically thick at wavelengths shorter than 130~\micron\ and has
associated class~I methanol masers \citep{CFGH78,KPFGW95,KS98}.
NGC6334I:IRS I2 is a deeply embedded mid-infrared source approximately
6 arcsec to the northwest of NGC6334I/NGC6334F.  It is much stronger
at 18 than 10~\micron\ with a dust temperature of about 110~K
\citep{DRPT02}.  The astrometry of \citet{DRPT02} find the methanol
masers to be projected against mid-infrared emission, but offset by
more than an arcsecond from the peak.  They suggest that the methanol
masers may instead be associated with a secondary peak in the
\ammonia(3,3) emission \citep{KPFGW95}.  In either scenario the
methanol masers are associated with a very young source.

NGC6334F is the brightest \ionhy\ region in our sample and
Fig.~\ref{fig:ngc6334f} shows significant extension in both the 6-km
and 750-m array images.  The dynamic range of images of this region
are limited by the presence of nearby more extended \ionhy\ regions.
NGC6334E can be seen in Fig.~\ref{fig:ngc6334f}b offset to the
north-west and there are signs of NGC6334B to the south-west, which is
clearly seen in the MOST image (Fig.~\ref{fig:ngc6334f}c).
Fig.~\ref{fig:ngc6334f}b shows that NGC6334E has an angular size of
approximately 50 arcseconds, significantly more than the 20 arcsec
estimate of \citet{RCM82} and demonstrates that as expected our 750-m
array observations are able to image emission on this scale.  The
location and size of E are well matched to a void in the dust and
molecular emission in the NGC6334 region \citep{S00,MSMKSDS00}, these
have presumably been destroyed and driven away by UV photons and
stellar winds.

\begin{figure}
\vsfig{ngc6334f_h}{a}
\vsfig{ngc6334f_m}{b}
\vsfig{ngc6334f_l}{c}
\caption{Radio continuum images of NGC6334F (a) 8.59-GHz ATCA 6-km image 
  of the compact emission \citep{ENM96}(b) 8.64-GHz ATCA 750-m image
  of the extended emission (c) 843-MHz MOST image of the large scale
  emission \citep{GCLY99}.  The location of the 6.7-GHz methanol maser
  clusters are marked with plus symbols.  The squares in (b) and (c)
  show the area covered in figures (a) and (b) respectively.}
\label{fig:ngc6334f}
\end{figure}

\section{Discussion} \label{sec:discussion}

Comparison of Fig.~\ref{fig:g308}-\ref{fig:ngc6334f} with comparable
figures in \citet{KWHO99} qualitatively shows that degree of extended
emission in our sample is in general much less.  The 6-km and 750-m
ATCA observations were made at the same frequency and have comparable
sensitivity in both scale size and intensity to the B- and D-array VLA
observations of \citet{KWHO99} and so the differences cannot be
attributed to observational effects.  Good quantitative measures of
the degree of extended emission are difficult, due to the often
complex morphology of the extended emission.  One simple means is to
compare the peak and integrated intensity and the relative intensity
between observations made in different array configurations.  For a
point source the peak flux measured in mJy beam$^{-1}$ will be equal
to the integrated flux density measured in mJy. So the amount by which
the integrated flux density exceeds the peak flux is a measure of the
percentage of emission present on scales larger than the synthesised
beam of the observations.  Similarly, a comparison of the integrated
intensity of the 6-km and 750-m array observations gives a direct
measure of the percentage of emission resolved out by the higher
resolution observations.  These quantities are summarised in
Table~\ref{tab:dynamic}.  The integrated flux reported in the table is
from a boxed region that encompasses the emission to the level of the
lowest contour shown in the figures.

\begin{table*}
  \caption{The peak flux density, integrated flux and RMS noise level in the 
    residual image for 11 \ionhy\ regions.  Sources indicated with a $^*$ do
    not have an associated 6.7-GHz methanol maser.  There is no emission 
    sufficiently compact to be imaged in the 6km array associated with 
    NGC6334E.}
  \begin{tabular}{lrrrrrrr} \hline
    {\bf Source} & \multicolumn{2}{c}{{\bf 6km Array Observations}} & 
    \multicolumn{2}{c}{{\bf 750m Array Observations}} & 
    \multicolumn{2}{c}{\bf Ratio of} \\
    & \multicolumn{1}{c}{{\bf Peak Flux}} & 
      \multicolumn{1}{c}{{\bf Integrated}} & 
      \multicolumn{1}{c}{{\bf Peak Flux}}  & 
      \multicolumn{1}{c}{{\bf Integrated}} & 
      \multicolumn{1}{c}{\bf 750m Int. to} & 
      \multicolumn{1}{c}{\bf 750m Int. to} \\
    & \multicolumn{1}{c}{{\bf Density}}    & 
      \multicolumn{1}{c}{{\bf Flux}} & 
      \multicolumn{1}{c}{{\bf Density}} & 
      \multicolumn{1}{c}{{\bf Flux}} & 
      \multicolumn{1}{c}{\bf 750m Peak} & 
      \multicolumn{1}{c}{\bf 6km Int.} \\
    & \multicolumn{1}{c}{{\bf (mJy beam$^{-1}$)}} & 
      \multicolumn{1}{c}{{\bf (mJy)}} & 
      \multicolumn{1}{c}{{\bf (mJy beam$^{-1}$)}} & 
      \multicolumn{1}{c}{{\bf (mJy)}} & & \\ \hline
    G\,308.92+0.12     &  25.0 &  363 &  263 &  488 & 1.86 & 1.34 \\ 
    G\,309.92+0.48     & 320.9 &  679 &  671 &  730 & 1.09 & 1.08 \\ 
    G\,318.91-0.16$^*$ &  40.8 & 1078 &  725 & 1513 & 2.09 & 1.40 \\ 
    G\,328.81+0.63     & 239.2 & 1680 & 1461 & 2033 & 1.39 & 1.21 \\ 
    G\,336.41-0.26     &   7.4 &   35 &   28 &   39 & 1.39 & 1.11 \\ 
    G\,339.88-1.26     &   7.6 &   10 &  6.6 &   22 & 3.33 & 2.20 \\ 
    G\,345.01+1.79     & 232.1 &  273 &  286 &  298 & 1.05 & 1.09 \\ 
    G\,345.00+1.79$^*$ &  7.3  &   62 &   60 &  128 & 2.13 & 2.06 \\
    G\,345.01+1.82$^*$ & 54.6  &  139 &  145 &  176 & 1.21 & 1.26 \\
    NGC6334F           & 479.6 & 2694 & 2130 & 4622 & 2.17 & 1.71 \\ 
    NGC6334E$^*$       &       &      &  417 & 5100 & 12.23 & \\ \hline
  \end{tabular}
  \label{tab:dynamic}
\end{table*}

\begin{figure}
  \psfig{file=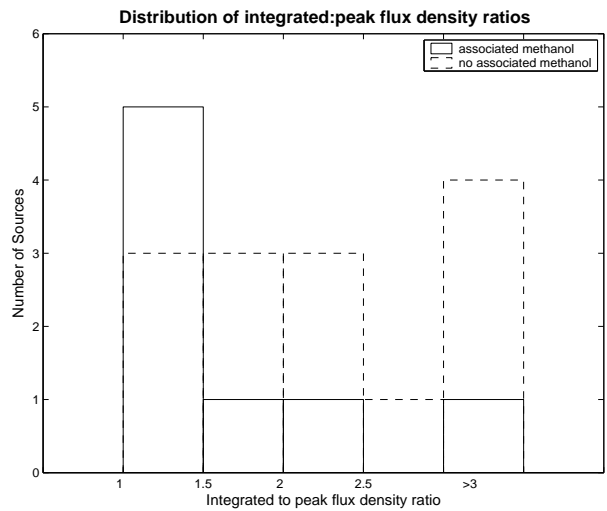,width=8cm}
  \caption{The distribution of integrated to peak flux density ratios
    for 750-m/D-Array observations.  The solid line histogram shows
    the distribution for radio continuum regions with associated
    class~II methanol masers and the dashed line shows the
    distribution for radio continuum regions without associated
    methanol masers.}
  \label{fig:histo}
\end{figure}

A direct quantitative comparison of the amount of extended emission in
\ionhy\ regions with and without associated class~II methanol masers
is shown in Fig.~\ref{fig:histo}.  A sample of 22 sources has been
compiled, the 11 reported in the current work and the 11 from
\citet{KWHO99} for which emission was detected in the D-array
observations.  We have not included the observations of \citet{KK01}
in this analysis as they are significantly different from those we
report here and those of \citeauthor{KWHO99}.  For the current
observations we have calculated the ratio of the integrated to the
peak flux density for each source detected in the 750-m array
observations, while for \citeauthor{KWHO99} we have taken the ratio of
the integrated flux density within a 50 arcsec square at the centre of
the field to the peak flux density for the D-array observations
(columns 5 \& 4 of their Table~4).  There are a total of eight sources
in the sample with associated methanol masers, the seven from the
current work, plus {\em IRAS}\ 18496+0004 from \citeauthor{KWHO99}.
{\em IRAS}\ 22543+6145/Cep A from the \citeauthor{KWHO99} list also
has an associated 6.7-GHz methanol maser maser, but no emission is
reported in their D-array observations.  There are fourteen sources in
the sample without associated methanol masers, the ten sources for
which \citet{KWHO99} list D-array observations and four sources from
the current work (marked with an asterisk in Table~\ref{tab:dynamic}).
Although the sample size is small, there is a clear tendency for
sources with associated methanol masers to have quantitatively less
extended emission.  Three of the eight sources (38 per cent) with
associated 6.7-GHz methanol masers have a ratio of integrated to peak
flux density greater than 1.5 (G\,308.92+0.12, G\,339.88-1.26 and
NGC6334F).  In contrast eleven of the fourteen sources without an
associated 6.7-GHz methanol maser (79 per cent) have a radio greater
than 1.5.  High-mass stars form in clusters and so we expect that for
some regions the observed radio continuum emission will be due to more
than one ionizing source, possibly at different evolutionary stages.
Considering that this will confuse the simple scenario we have
outlined and dilute the difference between the two samples, the
difference we find is striking.

A fundamental issue relating to \ionhy\ regions with both compact
components and extended emission is whether they are truly associated,
that is are they both produced by a single exciting star?
\citet{KWHO99} argued on the basis of morphology that in the majority
of cases where they see extended emission it is associated with the
compact component.  \citet{KK01} made single dish recombination line
observations towards their sample and found that in all but one case
both the compact and extended components have the same approximate
velocity.  For our sample of \ionhy\ regions associated with methanol
masers the limited degree and morphology of the extended emission
observed in most cases strongly suggests that it is associated with
the compact region.  This is confirmed by \citet{SEKF04} which used
the ATCA in the 750D array to make recombination line observations of
the same sample of \ionhy\ regions.  Recombination lines were detected
towards five of the eight sources and in each case the moment maps
demonstrate that the compact and extended emission are associated.

Our results and those of \citet{SEKF04} and \citet{KWHO99} appear to
be consistent with the scenario outlined in the introduction.  This
scenario is appealing as it explains both the lifetime problem and
association between compact and extended emission.  It is also
consistent with the observation that the majority of {\em IRAS}
sources with colours consistent with \UCHII\ regions do not have
associated 6.7-GHz methanol masers \citep{SK00}.  Our observations
suggest that those \ionhy\ regions with associated 6.7-GHz masers are
the young ones, while those without are likely to exhibit significant
extended radio continuum emission in addition to any compact
components.  The remaining hurdle for our scenario is a plausible
mechanism through which it can occur.  \citet{KK01} suggested that the
association between compact and extended radio continuum emission in
\ionhy\ regions may be due to champagne flows in a hierarchically
structure molecular cloud.  \citet{SEKF04} have used recombination
line observations and information from the literature to derive a
number of physical parameters for the \ionhy\ regions in this sample.
They have modelled the evolution of \ionhy\ regions in a hierarchical
molecular cloud formation and find good agreement between the observed
and predicted radii and emission measures.

There is a well known tendency for the stellar type as estimated from
the IR luminosity (typically {\em IRAS} observations) to exceed that
obtained using the radio flux density.  A variety of explanations have
been forwarded for this, including that at the spatial resolution of
{\em IRAS} the IR luminosity measured is that for the cluster, rather
than for an individual star and that dust absorbs some of the UV flux
from the star, hence reducing the radio flux density.  Our
observations and those of \citet{KWHO99} and \citet{KK01} suggest that
another factor in the discrepancy between IR and radio determined
spectral types is that most high resolution interferometry
observations significantly underestimate the total radio flux density
from the \ionhy\ regions due to their insensitivity to the extended
component.  This effect will be greatest for older \ionhy\ regions
where the fraction of the total flux density in the extended state
becomes a significant.

\section{Conclusions}

Observations of eight \ionhy\ regions associated with 6.7-GHz methanol
masers find a significantly lower degree of extended emission
associated with these sources than in other samples.  We suggest that
this is consistent with a scenario where both compact and diffuse
ionized structures co-exist for a significant fraction of \ionhy\ 
region lifetimes.  Modelling by \citet{SEKF04} of high-mass stars
forming in hierarchically structured molecular clouds is consistent
with our observations and appears to provide a consistent solution for
the lifetime problem.

\section*{Acknowledgements}
We would like to thank Marco Costa for his assistance with some of the
observations presented in this paper.  This research has made use of
NASA's Astrophysics Data System Abstract Service.  Financial support
for this work was provided by the Australian Research Council.  This
research has made use of the NASA/ IPAC Infrared Science Archive,
which is operated by the Jet Propulsion Laboratory, California
Institute of Technology, under contract with the National Aeronautics
and Space Administration.

\end{document}